\documentclass[12pt]{article}
\begin{document}
\newcommand{\beq}{\begin{equation}}
\newcommand{\eeq}{\end{equation}}
\newcommand{\beqa}{\begin{eqnarray}}
\newcommand{\eeqa}{\end{eqnarray}}
\newcommand{\beqar}{\begin{eqnarray*}}
\newcommand{\eeqar}{\end{eqnarray*}}
\newcommand{\al}{\alpha}
\newcommand{\be}{\beta}
\newcommand{\del}{\delta}
\newcommand{\D}{\Delta}
\newcommand{\eps}{\epsilon}
\newcommand{\ga}{\gamma}
\newcommand{\Ga}{\Gamma}
\newcommand{\ka}{\kappa}
\newcommand{\nn}{\nonumber}
\newcommand{\inn}{\!\cdot\!}
\newcommand{\h}{\eta}
\newcommand{\ii}{\iota}
\newcommand{\kk}{\varphi}
\newcommand\F{{}_3F_2}
\newcommand{\la}{\lambda}
\newcommand{\La}{\Lambda}
\newcommand{\na}{\prt}
\newcommand{\Om}{\Omega}
\newcommand{\om}{\omega}
\newcommand{\p}{\phi}
\newcommand{\sig}{\sigma}
\renewcommand{\t}{\theta}
\newcommand{\z}{\zeta}
\newcommand{\ssc}{\scriptscriptstyle}
\newcommand{\eg}{{\it e.g.,}\ }
\newcommand{\ie}{{\it i.e.,}\ }
\newcommand{\labell}[1]{\label{#1}} 
\newcommand{\reef}[1]{(\ref{#1})}
\newcommand\prt{\partial}
\newcommand\veps{\varepsilon}
\newcommand{\pol}{\varepsilon}
\newcommand\vp{\varphi}
\newcommand\ls{\ell_s}
\newcommand\dS{\dot{\cal S}}
\newcommand\dB{\dot{B}}
\newcommand\dG{\dot{G}}
\newcommand\ddG{\dot{\dot{G}}}
\newcommand\dP{\dot{\Phi}}
\newcommand\cF{{\cal F}}
\newcommand\cA{{\cal A}}
\newcommand\cS{{\cal S}}
\newcommand\cT{{\cal T}}
\newcommand\cV{{\cal V}}
\newcommand\cL{{\cal L}}
\newcommand\cM{{\cal M}}
\newcommand\cN{{\cal N}}
\newcommand\cG{{\cal G}}
\newcommand\cH{{\cal H}}
\newcommand\cI{{\cal I}}
\newcommand\cJ{{\cal J}}
\newcommand\cl{{\iota}}
\newcommand\cP{{\cal P}}
\newcommand\cQ{{\cal Q}}
\newcommand\cg{{\it g}}
\newcommand\cR{{\cal R}}
\newcommand\cB{{\cal B}}
\newcommand\cO{{\cal O}}
\newcommand\tcO{{\tilde {{\cal O}}}}
\newcommand\bg{\bar{g}}
\newcommand\bb{\bar{b}}
\newcommand\bH{\bar{H}}
\newcommand\bX{\bar{X}}
\newcommand\bK{\bar{K}}
\newcommand\bA{\bar{A}}
\newcommand\bZ{\bar{Z}}
\newcommand\bxi{\bar{\xi}}
\newcommand\bphi{\bar{\phi}}
\newcommand\bpsi{\bar{\psi}}
\newcommand\bprt{\bar{\prt}}
\newcommand\bet{\bar{\eta}}
\newcommand\btau{\bar{\tau}}
\newcommand\bnabla{\bar{\nabla}}
\newcommand\hF{\hat{F}}
\newcommand\hA{\hat{A}}
\newcommand\hT{\hat{T}}
\newcommand\htau{\hat{\tau}}
\newcommand\hD{\hat{D}}
\newcommand\hf{\hat{f}}
\newcommand\hg{\hat{g}}
\newcommand\hp{\hat{\phi}}
\newcommand\hi{\hat{i}}
\newcommand\ha{\hat{a}}
\newcommand\hb{\hat{b}}
\newcommand\hQ{\hat{Q}}
\newcommand\hP{\hat{\Phi}}
\newcommand\hS{\hat{S}}
\newcommand\hX{\hat{X}}
\newcommand\tL{\tilde{\cal L}}
\newcommand\hL{\hat{\cal L}}
\newcommand\tG{{\widetilde G}}
\newcommand\tg{{\widetilde g}}
\newcommand\tphi{{\widetilde \phi}}
\newcommand\tPhi{{\widetilde \Phi}}
\newcommand\td{{\tilde d}}
\newcommand\tk{{\tilde k}}
\newcommand\tf{{\tilde f}}
\newcommand\ta{{\tilde a}}
\newcommand\tb{{\tilde b}}
\newcommand\tc{{\tilde c}}
\newcommand\tR{{\tilde R}}
\newcommand\teta{{\tilde \eta}}
\newcommand\tF{{\widetilde F}}
\newcommand\tK{{\widetilde K}}
\newcommand\tE{{\widetilde E}}
\newcommand\tpsi{{\tilde \psi}}
\newcommand\tX{{\widetilde X}}
\newcommand\tD{{\widetilde D}}
\newcommand\tO{{\widetilde O}}
\newcommand\tS{{\tilde S}}
\newcommand\tB{{\widetilde B}}
\newcommand\tA{{\widetilde A}}
\newcommand\tT{{\widetilde T}}
\newcommand\tC{{\widetilde C}}
\newcommand\tV{{\widetilde V}}
\newcommand\thF{{\widetilde {\hat {F}}}}
\newcommand\Tr{{\rm Tr}}
\newcommand\tr{{\rm tr}}
\newcommand\STr{{\rm STr}}
\newcommand\hR{\hat{R}}
\newcommand\M[2]{M^{#1}{}_{#2}}

\newcommand\bS{\textbf{ S}}
\newcommand\bI{\textbf{ I}}
\newcommand\bJ{\textbf{ J}}

\begin{titlepage}
\begin{center}

\vskip 2 cm
{\LARGE \bf  $O(25,25)$ symmetry of bosonic string theory  \\ \vskip 0.75  cm   at order $\alpha'^2$     }\\
\vskip 1.25 cm
   Mohammad R. Garousi\footnote{garousi@um.ac.ir}

\vskip 1 cm
{{\it Department of Physics, Faculty of Science, Ferdowsi University of Mashhad\\}{\it P.O. Box 1436, Mashhad, Iran}\\}
\vskip .1 cm
 \end{center}

\begin{abstract}
It has been   recently observed that the imposition of the $O(1,1)$ symmetry on  the circle reduction of  the  classical  effective action  of string theory,  can fix the  effective action of the bosonic string theory  at order $\alpha'^2$, up to an overall factor.  In this paper, we use the cosmological reduction on the action and show that, up to one-dimensional field redefinitions and total derivative terms, it can be written in   the  $O(25,25)$-invariant form proposed  by Hohm and Zwiebach. 
\end{abstract}
\end{titlepage}

\section{Introduction}

The string theory is a   consistent theory of quantum  gravity  which includes  the finite number of massless modes and the tower of infinite number of  massive modes of the string excitations. At the low energies,   the massive modes are integrated out to produce a higher-derivative effective action which includes only the massless fields.  The effective action then   has the genus-expansion  and the   stringy-expansion which is an expansion in terms of  higher derivative couplings  at each loop level.  The classical    effective action of the bosonic string theory on a spacetime manifold which has no boundary, has the following  higher-derivative or  $\alpha'$-expansion:
\beqa
\bS_{\rm eff}&=&\sum^\infty_{m=0}\alpha'^m\bS_m=\bS_0+\alpha' \bS_1 +\alpha'^2 \bS_2+\alpha'^3 \bS_3+\cdots \labell{seff}
\eeqa 
The leading order  action $\!\!\bS_0$ is the gauge invariant  two-derivative couplings of the massless fields which includes, among other things, the Hilbert-Einstein term.  This action and their appropriate  higher derivative extensions may be found by the  S-matrix method \cite{Gross:1986iv,Gross:1986mw}, by  the sigma-mode method \cite{Grisaru:1986vi,Freeman:1986zh} or by exploring  various symmetries in  the  string theory.

One of the symmetries  in the perturbative  string theory is   T-duality \cite{Giveon:1994fu,Alvarez:1994dn}  which appears  when one compactifies theory  on a torus, \eg the compactification of the full bosonic string theory on tours $T^d$ is invariant under   $O(d,d,Z)$ transformations. After integrating out the massive modes, however, the T-duality  should appears as symmetry in the  effective actions. It has been proved in \cite{Sen:1991zi} that the dimensional reduction of the classical effective actions of the bosonic  string theory at each order of $\alpha'$  is  invariant under  $O(d,d,R)$ transformations. 

In the most simple case, when one reduces the effective action  on a circle, the invariance of the reduced action under the $Z_2$-subgroup of the $O(1,1)$-group constrains greatly the  couplings in the  effective action. This constraint and the constraint that the couplings in the effective action must be invariant under the gauge transformations, fix the couplings at the leading order of $\alpha'$, \ie
\beqa
\bS_0
=-\frac{2}{\kappa^2} \int d^{26}x \, e^{-2\phi}\sqrt{-G} \left(  R + 4\nabla_{\mu}\phi \nabla^{\mu}\phi-\frac{1}{12}H^2\right)\labell{action1}
\eeqa
which is the standard effective action of the bosonic string theory. These constraints  fix the 
 couplings at order $\alpha', \alpha'^2$ up to an overall factor \cite{Garousi:2019wgz,Garousi:2019mca}, \ie
  \beqa
 \bS_1&=&\frac{-2b_1 }{\kappa^2}\int d^{26}x\, e^{-2\phi}\sqrt{-G}\Big(   R_{\alpha \beta \gamma \delta} R^{\alpha \beta \gamma \delta} -\frac{1}{2}H_{\alpha}{}^{\delta \epsilon} H^{\alpha \beta \gamma} R_{\beta  \gamma \delta\epsilon}\nn\\
&&\qquad\qquad\qquad\qquad\qquad\quad+\frac{1}{24}H_{\epsilon\delta \zeta}H^{\epsilon}{}_{\alpha}{}^{\beta}H^{\delta}{}_{\beta}{}^{\gamma}H^{\zeta}{}_{\gamma}{}^{\alpha}-\frac{1}{8}H_{\alpha \beta}{}^{\delta} H^{\alpha \beta \gamma} H_{\gamma}{}^{\epsilon \zeta} H_{\delta \epsilon \zeta}\Big)\labell{S1bf}
\eeqa
\beqa
\bS_2&=&\frac{-2b_1^2 }{\kappa^2}\int d^{26}x \,e^{-2\phi}\sqrt{-G}\Big(  - \frac{1}{12} H_{\alpha}{}^{\delta \epsilon} H^{\alpha \beta 
\gamma} H_{\beta \delta}{}^{\zeta} H_{\gamma}{}^{\iota \kappa} 
H_{\epsilon \iota}{}^{\mu} H_{\zeta \kappa \mu}\nn\\&& + 
\frac{1}{30} H_{\alpha \beta}{}^{\delta} H^{\alpha \beta 
\gamma} H_{\gamma}{}^{\epsilon \zeta} H_{\delta}{}^{\iota 
\kappa} H_{\epsilon \zeta}{}^{\mu} H_{\iota \kappa \mu} + 
\frac{3}{10} H_{\alpha \beta}{}^{\delta} H^{\alpha \beta 
\gamma} H_{\gamma}{}^{\epsilon \zeta} H_{\delta 
\epsilon}{}^{\iota} H_{\zeta}{}^{\kappa \mu} H_{\iota \kappa 
\mu} \nn\\&&+ \frac{13}{20} H_{\alpha}{}^{\epsilon \zeta} 
H_{\beta}{}^{\iota \kappa} H_{\gamma \epsilon \zeta} H_{\delta 
\iota \kappa} R^{\alpha \beta \gamma \delta} + \frac{2}{5} 
H_{\alpha}{}^{\epsilon \zeta} H_{\beta \epsilon}{}^{\iota} 
H_{\gamma \zeta}{}^{\kappa} H_{\delta \iota \kappa} R^{\alpha 
\beta \gamma \delta}\nn\\&& + \frac{18}{5} H_{\alpha 
\gamma}{}^{\epsilon} H_{\beta}{}^{\zeta \iota} H_{\delta 
\zeta}{}^{\kappa} H_{\epsilon \iota \kappa} R^{\alpha \beta 
\gamma \delta} -  \frac{43}{5} H_{\alpha \gamma}{}^{\epsilon} 
H_{\beta}{}^{\zeta \iota} H_{\delta \epsilon}{}^{\kappa} 
H_{\zeta \iota \kappa} R^{\alpha \beta \gamma \delta} \nn\\&&-  
\frac{16}{5} H_{\alpha \gamma}{}^{\epsilon} H_{\beta 
\delta}{}^{\zeta} H_{\epsilon}{}^{\iota \kappa} H_{\zeta \iota 
\kappa} R^{\alpha \beta \gamma \delta} - 2 H_{\beta 
\epsilon}{}^{\iota} H_{\delta \zeta \iota} 
R_{\alpha}{}^{\epsilon}{}_{\gamma}{}^{\zeta} R^{\alpha \beta 
\gamma \delta} - 2 H_{\beta \delta}{}^{\iota} H_{\epsilon 
\zeta \iota} R_{\alpha}{}^{\epsilon}{}_{\gamma}{}^{\zeta} 
R^{\alpha \beta \gamma \delta}\nn\\&& -  \frac{4}{3} 
R_{\alpha}{}^{\epsilon}{}_{\gamma}{}^{\zeta} R^{\alpha \beta 
\gamma \delta} R_{\beta \zeta \delta \epsilon} + \frac{4}{3} 
R_{\alpha \beta}{}^{\epsilon \zeta} R^{\alpha \beta \gamma 
\delta} R_{\gamma \epsilon \delta \zeta} + 3 
H_{\beta}{}^{\zeta \iota} H_{\epsilon \zeta \iota} R^{\alpha 
\beta \gamma \delta} R_{\gamma}{}^{\epsilon}{}_{\alpha \delta} 
\nn\\&&+ 2 H_{\beta \epsilon}{}^{\iota} H_{\delta \zeta \iota} 
R^{\alpha \beta \gamma \delta} 
R_{\gamma}{}^{\epsilon}{}_{\alpha}{}^{\zeta} + 2 H_{\alpha 
\beta \epsilon} H_{\delta \zeta \iota} R^{\alpha \beta \gamma 
\delta} R_{\gamma}{}^{\epsilon \zeta \iota} + \frac{13}{10} 
H_{\alpha}{}^{\gamma \delta} H_{\beta \gamma}{}^{\epsilon} 
H_{\delta}{}^{\zeta \iota} H_{\epsilon \zeta \iota} 
\nabla^{\beta}\nabla^{\alpha}\phi\nn\\&& + \frac{13}{5} 
H_{\gamma}{}^{\epsilon \zeta} H_{\delta \epsilon \zeta} 
R_{\alpha}{}^{\gamma}{}_{\beta}{}^{\delta} 
\nabla^{\beta}\nabla^{\alpha}\phi -  \frac{52}{5} H_{\beta 
\delta}{}^{\zeta} H_{\gamma \epsilon \zeta} 
R_{\alpha}{}^{\gamma \delta \epsilon} 
\nabla^{\beta}\nabla^{\alpha}\phi\nn\\&& -  \frac{26}{5} H_{\alpha 
\gamma \epsilon} H_{\beta \delta \zeta} R^{\gamma \delta 
\epsilon \zeta} \nabla^{\beta}\nabla^{\alpha}\phi + 
\frac{13}{5} \nabla^{\beta}\nabla^{\alpha}\phi 
\nabla_{\epsilon}H_{\beta \gamma \delta} 
\nabla^{\epsilon}H_{\alpha}{}^{\gamma \delta} \nn\\&&+ \frac{13}{10} 
H_{\beta \gamma}{}^{\epsilon} H^{\beta \gamma \delta} 
H_{\delta}{}^{\zeta \iota} \nabla^{\alpha}\phi 
\nabla_{\iota}H_{\alpha \epsilon \zeta}  -  \frac{13}{20} 
H_{\alpha}{}^{\beta \gamma} H_{\delta \epsilon}{}^{\iota} 
H^{\delta \epsilon \zeta} \nabla^{\alpha}\phi 
\nabla_{\iota}H_{\beta \gamma \zeta} \nn\\&&+ \frac{1}{20} 
H_{\alpha}{}^{\delta \epsilon} H^{\alpha \beta \gamma} 
\nabla_{\iota}H_{\delta \epsilon \zeta} 
\nabla^{\iota}H_{\beta \gamma}{}^{\zeta} + \frac{1}{5} 
H_{\alpha}{}^{\delta \epsilon} H^{\alpha \beta \gamma} 
\nabla_{\zeta}H_{\gamma \epsilon \iota} 
\nabla^{\iota}H_{\beta \delta}{}^{\zeta}\nn\\&& -  \frac{6}{5} 
H_{\alpha}{}^{\delta \epsilon} H^{\alpha \beta \gamma} 
\nabla_{\iota}H_{\gamma \epsilon \zeta} 
\nabla^{\iota}H_{\beta \delta}{}^{\zeta} -  \frac{6}{5} 
H_{\alpha \beta}{}^{\delta} H^{\alpha \beta \gamma} 
\nabla_{\zeta}H_{\delta \epsilon \iota} 
\nabla^{\iota}H_{\gamma}{}^{\epsilon \zeta}\nn\\&&\qquad\qquad\qquad\qquad\qquad\qquad\qquad  + \frac{17}{10} 
H_{\alpha \beta}{}^{\delta} H^{\alpha \beta \gamma} 
\nabla_{\iota}H_{\delta \epsilon \zeta} 
\nabla^{\iota}H_{\gamma}{}^{\epsilon \zeta}\Big)\labell{S2f}
\eeqa
These constraints have been also used in \cite{Garousi:2020gio,Garousi:2020lof} to find the NS-NS couplings in type II superstring theory at order $\alpha'^3$ up to an overall factor. The  couplings  \reef{S1bf}, for  the overall factor $b_1=1/4$,  are   the standard effective action of the  bosonic string theory at order $\alpha'$  which has been found in \cite{Metsaev:1987zx} by the S-matrix calculations\footnote{Note that there is a typo in the overall coeffient of $\bS_2$ in \cite{Garousi:2019mca}, \ie the factor $b_1$ in \cite{Garousi:2019mca} should be $b_1^2$.}.  The two Riemann cubed terms in \reef{S2f} are  the gravity couplings  that have been found in  \cite{Metsaev:1986yb} by the S-matrix method. Using field redefinitions, the above couplings can be rewritten in many other schemes. The $B$-field couplings in \reef{S2f} can be reproduced by the S-matrix method \cite{Gholian}. In this paper, we are going to check these couplings  by studying the  cosmological reduction of the couplings and show that they are invariant under $O(25,25)$ transformations. Similar calculation has been done in \cite{Codina:2020kvj,Garousi:2021ikb} for the NS-NS couplings in the  type II superstring theory at order $\alpha'^3$.

When one uses the cosmological reduction on the classical effective action, the resulting one-dimensional effective action should have $O(d,d,R)$ symmetry \cite{Sen:1991zi,Hohm:2014sxa}.
Using the most general one-dimensional field redefinitions, including  the lapse function, and  using  integration by part, it has been shown in \cite{Hohm:2015doa,Hohm:2019jgu} that the cosmological reduction of the effective action \reef{seff} at order $\alpha'$ and higher, can be written in a scheme in which only  the first time-derivative of the generalized metric ${\cal S}$ appears. Trace of odd number of the first-derivative of $\cS$  is zero. It has been shown in \cite{Hohm:2019jgu} that the couplings which include  $\tr(\dS^2)$ can be removed by the lapse function transformation. Then the one-dimensional  action can be written in a minimal scheme as  the following expansion \cite{Hohm:2015doa,Hohm:2019jgu}:
\beqa
S_{\rm eff}^c&=&S_0^c+\int dt e^{-\Phi}\bigg(\alpha' c_{2,0}\tr(\dS^4)+\alpha'^2c_{3,0}\tr(\dS^6)\nn\\
&&\qquad\quad\qquad\quad+\alpha'^3[c_{4,0}\tr(\dS^8)+c_{4,1}(\tr(\dS^4))^2]\nn\\
&&\qquad\quad\qquad\quad+\alpha'^4[c_{5,0}\tr(\dS^{10})+c_{5,1}\tr(\dS^6)\tr(\dS^4)]+\cdots\bigg)\labell{cosm}
\eeqa
where the coefficients $c_{m,n}$ depends on the specific theory, \eg $c_{2,0}, c_{3,0}$ are non-zero for the bosonic string theory whereas these numbers are zero for the superstring theory. The lapse function in above action is set $n=1$. The de Sitter vacua of this action  have been studied in \cite{Hohm:2019ccp,Nunez:2020hxx,Basile:2021krk}.

The above minimal action has been found in \cite{Hohm:2015doa,Hohm:2019jgu} by imposing various equations of motion which is the same as  using field redefinitions and  removing total derivative terms.  We are going to show that by imposing the most general field redefinitions and total derivative terms, the cosmological reduction of the actions \reef{S1bf} and \reef{S2f} can be written in the above form. In this way, by keeping track the field redefinitions and total derivative terms, one can study the cosmological reduction of the boundary actions \cite{Garousi:2019xlf,Garousi:2021cfc} as well, in which we are not interested in this paper.

Under the field redefinition $\psi\rightarrow \psi+\alpha'\delta\psi^{(1)}+\alpha'^2\delta\psi^{(2)}$ where $\psi$  collectively represents the one-dimensional functions, the cosmological reduction of the actions in \reef{seff}  have the following expansions:
\beqa
\bS_0^c(\psi+\alpha'\delta\psi^{(1)}+\alpha'^2\delta\psi^{(2)})&=&\bS_0^c(\psi)+\alpha'\delta_1\!\!\bS_0^c(\psi)+\frac{1}{2}\alpha'^2\delta_1^2\!\!\bS_0^c(\psi)+\alpha'^2\delta_2\!\!\bS_0^c(\psi)+\cdots\nn\\
\alpha'\!\!\bS_1^c(\psi+\alpha'\delta\psi^{(1)}+\alpha'^2\delta\psi^{(2)})&=&\alpha'\!\!\bS_1^c(\psi)+\alpha'^2\delta_1\!\!\bS_1^c(\psi)+\cdots\nn\\
\alpha'^2\!\!\bS_2^c(\psi+\alpha'\delta\psi^{(1)}+\alpha'^2\delta\psi^{(2)})&=&\alpha'^2\!\!\bS_2^c(\psi)+\cdots
\eeqa
where dots represent term at order $\alpha'^3$ and higher in which we are not interested. In above equations, $\delta_i$ on the action means the action contains the  first order perturbation $\delta\psi^{(i)}$ and   $\delta_1^2$ on the action means the action contains the  second order perturbation $\delta\psi^{(1)}\delta\psi^{(1)}$. Then, up to some total derivative terms, the sum of terms at each order of $\alpha'$, \ie 
\beqa
S_0^c&=&\!\!\bS_0^c(\psi)\nn\\
S_1^c&=&\!\!\bS_1^c(\psi)+\delta_1\!\!\bS_0^c(\psi)\labell{S012}\\
S_2^c&=&\!\!\bS_2^c(\psi)+\delta_1\!\!\bS_1^c(\psi)+\frac{1}{2}\delta_1^2\!\!\bS_0^c(\psi)+\delta_2\!\!\bS_0^c(\psi)\nn
\eeqa
 should be written in $O(25,25)$-invariant form as in  \reef{cosm}. To find the above perturbations, one needs to know how the lapse function $n$ appears in the cosmological action $\!\!\bS_0^c$, $\!\!\bS_1^c$, $\cdots$ because the field redefinition should include the redefinition of all functions including the laps function. In general, the derivatives of the lapse function may appear in some complicated expressions in  these actions. However, the $O(25,25)$-covariance forces this function to appear in the $O(25,25)$-invariant form of the actions  by replacing the measure of integral as $dt\rightarrow dt/n^{(m-1)}$ where  $2m$ is the number of time-derivative in the action, \eg the laps function in $S_1^c$ is as $dt\rightarrow dt/n^3$.

Since we know how the lapse function appears in the actions $S_0^c,S_1^c,S_2^2,\cdots$, it is convenient to use the field redefinitions in these actions, \ie
\beqa
S_0^c(\psi+\alpha'\delta\psi^{(1)}+\alpha'^2\delta\psi^{(2)})&=&S_0^c(\psi)+\alpha'\delta_1S_0^c(\psi)+\frac{1}{2}\alpha'^2\delta_1^2S_0^c(\psi)+\alpha'^2\delta_2S_0^c(\psi)+\cdots\nn\\
\alpha'S_1^c(\psi+\alpha'\delta_1\psi+\alpha'^2\delta_2\psi)&=&\alpha'S_1^c(\psi)+\alpha'^2\delta_1S_1^c(\psi)+\cdots\labell{S0121}\\
\alpha'^2S_2^c(\psi+\alpha'\delta_1\psi+\alpha'^2\delta_2\psi)&=&\alpha'^2S_2^c(\psi)+\cdots\nn
\eeqa
 where the perturbations  can easily be calculated, once we know the $O(25,25)$-invariant form of the actions. Hence, one has to relate the perturbations $\delta\!\!\bS_0^c,\delta\!\!\bS_1^2,\cdots$ in \reef{S012} to the perturbations  $\delta S_0^c,\delta S_1^c,\cdots$. It is obvious that  $\delta\!\!\bS_0^c=\delta S_0^c$ because the leading order action is $O(25,25)$ invariant up to a total derivative term. To find the relation between $\delta\!\!\bS_1^c$ and $\delta S_0^c,\delta S_1^c$, we expand the right-hand side of second equation in \reef{S012} and compare it with the right-hand side of the second equation in \reef{S0121} to find the following relation:
 \beqa
 \delta_1\!\!\bS_1^c&=&\delta_1 S_1^c-\delta_1^2  S_0^c
 \eeqa
Inserting them into \reef{S012}, one finds the following expressions should be invariant under the $O(25,25)$ transformations: 
\beqa
S_0^c&=&\!\!\bS_0^c(\psi)\nn\\
S_1^c&=&\!\!\bS_1^c(\psi)+\delta_1S_0^c(\psi)\labell{S0122}\\
S_2^c&=&\!\!\bS_2^c(\psi)+\delta_1S_1^c(\psi)-\frac{1}{2}\delta_1^2S_0^c(\psi)+\delta_2S_0^c(\psi)\nn
\eeqa
which contains only the perturbation of the $O(25,25)$-invariant actions. In this paper, we are going to show that when the cosmological reduction of the actions \reef{action1}, \reef{S1bf} and \reef{S2f} are inserted in the above expressions, they satisfy the $O(25,25)$ symmetry and can be written in the minimal form of \reef{cosm}.

The outline of the paper is as follows:  In section 2, we review the observation that the cosmological reduction of the  leading order action is invariant under the $O(25,25)$ transformations.   In section 3, we show that when the  cosmological reduction of the action \reef{S1bf} is inserted into the right-hand side of the second expression in \reef{S0122}, it can be written in the standard form of  \reef{cosm} after including  the appropriate one-dimensional field redefinition $\delta_1S_0^c(\psi)$ and total derivative terms. In section 4, we show that when the  cosmological reduction of the action \reef{S2f} are inserted into the last expression in \reef{S0122}, the left-hand side can be written in the $O(25,25)$-invariant form of \reef{cosm}. This observation not only confirms the proposal \reef{cosm}, but also it confirms the effective action \reef{S2f} that has been found in \cite{Garousi:2019mca} by imposition of $O(1,1)$ symmetry on the most general gauge invariant couplings.

\section{Cosmological reduction at the leading order}

In this section, we review the cosmological reduction of the leading order  action \cite{Veneziano:1991ek,Meissner:1991zj,Maharana:1992my,Meissner:1996sa}. 
When fields depend only on time, using the gauge symmetries it is possible to write the metric, $B$-field  and dilaton as
 \beqa
G_{\mu\nu}=\left(\matrix{-n^2(t)& 0&\cr 0&G_{ij}(t)&}\right),\, B_{\mu\nu}= \left(\matrix{0&0\cr0&B_{ij}(t)&}\right),\,  2\phi=\Phi+\frac{1}{2}\log\det(G_{ij})\labell{creduce}\eeqa
where the lapse function $n(t)$ can also be fixed to $n=1$. The cosmological reduction of the  action \reef{action1} then becomes
\beqa
\bS_0^c&=&-\frac{2}{\kappa^2}\int dt e^{-\Phi}\Bigg[\frac{1}{4}\dB_{ij}\dB^{ij}-\frac{3}{4}\dG_{ij}\dG^{ij}-G^{ij}\dG_{ij}\dP-\dP^2+G^{ij}\ddot{G}_{ij}\Bigg]
\eeqa
where $\dG^{ij}\equiv G^{ik}G^{il}\dG_{kl}$. Removing a  total derivative term,
one can write $\bS_0^c$ as
\beqa
S_0^c&=&-\frac{2}{\kappa^2}\int dt e^{-\Phi}\Bigg[\frac{1}{4}\dB_{ij}\dB^{ij}+\frac{1}{4}\dG_{ij}\dG^{ij}-\dP^2\Bigg]\labell{S0c}
\eeqa
Using the generalized metric $\cS$ which is defined as 
\beqa
\cS\equiv \eta \left(\matrix{G^{-1}& -G^{-1}B&\cr BG^{-1}&G-BG^{-1}B&}\right)\labell{S}
\eeqa
where $\eta$ is the  metric of the $O(25,25)$ group which in the non-diagonal form is 
\beqa
\eta&=& \left(\matrix{0& 1&\cr 1&0&}\right),
\eeqa
one can write the  action \reef{S0c} as
\beqa
S_0^c&=&-\frac{2}{\kappa^2}\int dt e^{-\Phi}\Bigg[-\dP^2-\frac{1}{8}\tr(\dS^2)\Bigg]\labell{S0}
\eeqa
which is invariant under the global $O(25,25)$ transformations because the one-dimensional dilaton is invariant and the generalized metric transforms as
\beqa
\cS&\rightarrow &\Omega^T\cS\Omega
\eeqa
where $\Omega$ belong to the  $O(25,25)$ group, \ie $\Omega^T\eta\Omega=\eta$. The laps function can be inserted in the  action \reef{S0} by replacing $dt\rightarrow dt/n$.
 
\section{Cosmological reduction at order $\alpha'$}

In this section, we are going to show that the cosmological reduction of the couplings \reef{S1bf} can be written in $O(25,25)$-invariant after using appropriate one-dimensional field redefinitions and total derivative terms.

To find the cosmological reduction of these couplings we first find the cosmological reduction of the Riemann curvature and $H$. They are
\beqa
&&R_{ijkl}=-\frac{1}{4}\dG_{il}\dG_{jk}+\frac{1}{4}\dG_{ik}\dG_{jl}\,\,;\,\, R_{i0jk}=0\,\,;\,\,R_{i0j0}=\frac{1}{4}\dG_{ik}\dG^k{}_j-\frac{1}{2}\ddot{G}_{ij}\nn\\
&&H_{ijk}=0\,\,;\,\, H_{ij0}=\dB_{ij}\labell{red1}
\eeqa
Using the above reductions,  one finds the following cosmological reduction for  the action \reef{S1bf}:
\beqa
 \bS_1^c&=&-\frac{2b_1}{\kappa^2}\int dt e^{-\Phi}\Bigg[- \frac{3}{8} \dB_{i}{}^{k} \dB^{ij} \dB_{j}{}^{l} \dB_{kl} -  
\frac{1}{8} \dB_{ij} \dB^{ij} \dB_{kl} \dB^{kl} + \frac{1}{4} 
\dB^{ij} \dB^{kl} \dG_{ik} \dG_{jl} \nn\\&&\quad\qquad\qquad\qquad-  \frac{1}{2} \dB_{i}{}^{k} 
\dB^{ij} \dG_{j}{}^{l} \dG_{kl} + \frac{1}{8} \dG_{i}{}^{k} 
\dG^{ij} \dG_{j}{}^{l} \dG_{kl} + \frac{1}{8} \dG_{ij} \dG^{ij} 
\dG_{kl} \dG^{kl}\nn\\&&\quad\qquad\qquad\qquad + \ddot{G}_{ij} \ddot{G}^{ij} + \dB_{i}{}^{k} \dB^{ij} 
\ddot{G}_{jk} -  \dG_{i}{}^{k} \dG^{ij} \ddot{G}_{jk}\Bigg]\labell{cS3}
 \eeqa
where we have used the gauge $n=1$. This action is not invariant under $O(25,25)$ transformations. Some of the non-invariant terms are total  derivative terms  which should be removed. Moreover, the action in terms of the variables $G_{ij}, B_{ij},\Phi$ are not invariant. It should  be invariant in terms of some other variables which involve higher derivatives of  $G_{ij}, B_{ij},\Phi$.

To remove the total derivatives terms  from \reef{cS3}, we  add all total derivative terms at order $\alpha'$  with arbitrary coefficients to \reef{cS3}. We add the following total derivative terms:
\beqa
-\frac{2}{\kappa^2}\int dt\frac{d}{dt}(e^{-\Phi}\cI_1)
\eeqa
where  $\cI_1$ is   all possible    terms at three-derivative level with even parity which are constructed from $\dP$, $\dB$, $\dG$, $\ddot{\Phi}$, $\ddot{B}$, $\ddot{G}$,  $\cdots$.  Using the package   "xAct" \cite{Nutma:2013zea}, one finds there are 18 such terms, \ie
\beqa
\cI_1&=&j_1(\dP)^3+\cdots
\eeqa
where the coefficients  $J_1,\cdots, J_{18}$ are 18 arbitrary parameters.

One can change the field variables in \reef{creduce} as 
\begin{eqnarray}
G_{ij}&\rightarrow &G_{ij}+\alpha' \delta G^{(1)}_{ij}\nn\\
B_{ij}&\rightarrow &B_{ij}+ \alpha'\delta B^{(1)}_{ij}\nn\\
\Phi &\rightarrow &\Phi+ \alpha'\delta\Phi^{(1)}\nn\\
n &\rightarrow &n+ \alpha' \delta n^{(1)}\labell{gbpn}
\end{eqnarray}
where the matrices  $\delta G^{(1)}_{ij}$, $\delta B^{(1)}_{ij}$ and $\delta\Phi^{(1)}, \delta n^{(1)}$ are all possible  terms at 2-derivative level constructed from $\dP$, $\dB$, $\dG$, $\ddot{\Phi}$, $\ddot{B}$, $\ddot{G}$.  The perturbations  $\delta G^{(1)}_{ij}$, $\delta\Phi^{(1)}$, $\delta n^{(1)}$ contain even-parity terms and $\delta B^{(1)}_{ij}$ contains odd-parity terms, \ie
\beqa
\delta n^{(1)}&=&n_1\dB_i{}^j\dB_j{}^i+\cdots\nn\\
\delta \Phi^{(1)}&=&e_1\dB_i{}^j\dB_j{}^i+\cdots\nn\\
\delta G^{(1)}_{ij}&=&d_1\dB_i{}^k\dB_k{}_j+\cdots\nn\\
\delta B^{(1)}_{ij}&=&f_1\dG_i{}^k\dB_k{}_j+\cdots\labell{per1}
\eeqa
The coefficients $n_1,\cdots, n_{7}$, $e_1,\cdots, e_{7}$, $d_1,\cdots, d_{12}$  and $f_1,\cdots, f_{4}$ are arbitrary parameters.  When the field variables in $S_0^c$  are changed according to the above  field redefinitions, they produce some couplings at order $\alpha'$ and higher. In this section we are interested in the resulting couplings at order $\alpha'$, \ie 
\beqa
\delta_1 S_0^c&=&-\frac{2}{\kappa^2}\int dt e^{-\Phi}\Bigg[\delta n^{(1)}\left(-\frac{1}{4}\dB_{ij}\dB^{ij}-\frac{1}{4}\dG_{ij}\dG^{ij}+\dP^2\right)\labell{dS0c}\\
&&+\delta \Phi^{(1)}\left(-\frac{1}{4}\dB_{ij}\dB^{ij}-\frac{1}{4}\dG_{ij}\dG^{ij}+\dP^2\right)-2\dP\frac{d}{dt}\delta \Phi^{(1)}\nn\\
&&+\delta G^{(1)}_{ij}\left(-\frac{1}{2}\dB_k{}^j\dB^{ki}-\frac{1}{2}\dG_k{}^j\dG^{ki}\right)+\frac{1}{2}\dG^{ij}\frac{d}{dt}\delta G^{(1)}_{ij}+\frac{1}{2}\dB^{ij}\frac{d}{dt}\delta B^{(1)}_{ij}\Bigg]\nn
\eeqa
where we have used the fact that the lapse function appears in the action \reef{S0c} by replacing $dt\rightarrow dt/n$. When replacing the perturbations \reef{per1} in the above equation, one finds, for some relations between the parameters in \reef{per1}, the  above equation produces  total derivative terms. They are not correspond to the field redefinitions. One should remove such parameters in \reef{per1} to find the independent parameters which correspond to the pure field redefinitions.

Using the following field redefinitions:
 \beqa
\delta n^{(1)}&=&b_1\Big(- \frac{1}{2} \dB_{ij} \dB^{ij} -  \frac{1}{2} \dP^2 + 2 \ddot{\Phi}\Big)\nn\\
\delta \Phi^{(1)}&=&b_1\Big(\frac{1}{2} \dG_{ij} \dG^{ij} + \frac{1}{2} \dP^2\Big)\nn\\
\delta G^{(1)}_{ij}&=&b_1\Big(- \dB_{i}{}^{k} \dB_{jk} - 3 \dG_{i}{}^{k} \dG_{jk} + 2 \ddot{G}_{ij}\Big)\nn\\
\delta B^{(1)}_{ij}&=&b_1\Big(\dB_{j}{}^{k} \dG_{ik} -  \dB_{i}{}^{k} \dG_{jk} + 2 \dB_{ij} 
\dP\Big) \labell{dG1dB1}
\eeqa
 one finds the cosmological action \reef{cS3} can be written as 
 \beqa
 S_1^c&=&\!\!\bS_1^c+\delta_1 S_0^c\,=\,-\frac{2b_1}{\kappa^2}\int dt e^{-\Phi}\Bigg[\frac{1}{8} \dB_{i}{}^{k} \dB^{ij} \dB_{j}{}^{l} \dB_{kl} -  
\frac{1}{4} \dB^{ij} \dB^{kl} \dG_{ik} \dG_{jl}\nn\\&&\qquad\qquad\qquad\quad + \frac{1}{2} 
\dB_{i}{}^{k} \dB^{ij} \dG_{j}{}^{l} \dG_{kl} + \frac{1}{8} 
\dG_{i}{}^{k} \dG^{ij} \dG_{j}{}^{l} \dG_{kl}\Bigg]\labell{action2}
 \eeqa
up to  the following total derivative terms:
\beqa
\cI_1&=&b_1\Big(\dB_{i}{}^{k} \dB^{ij} \dG_{jk} + \dG_{i}{}^{k} \dG^{ij} \dG_{jk} -  
\frac{1}{2} \dB_{ij} \dB^{ij} \dP + \frac{1}{2} \dG_{ij} 
\dG^{ij} \dP -  \dG^{ij} \ddot{G}_{ij}\Big)
\eeqa
The action \reef{action2}  has only the first time-derivative on matrices $G_{ij},B_{ij}$ and has no trace of one $\dG$ and two $\dG$ or $\dB$.
 
Now using the definition of the generalized metric in \reef{S}, one finds
\beqa
\tr(\dS^4)&=&2 \dB_{i}{}^{k} \dB^{ij} \dB_{j}{}^{l} \dB_{kl} - 4 \dB^{ij} 
\dB^{kl} \dG_{ik} \dG_{jl} + 8 \dB_{i}{}^{k} \dB^{ij} \dG_{j}{}^{l} 
\dG_{kl} + 2 \dG_{i}{}^{k} \dG^{ij} \dG_{j}{}^{l} \dG_{kl}
\eeqa
 Using the above $O(25,25)$-invariant expressions, one can write \reef{action2} as 
\beqa
 S_1^c&=&-\frac{2b_1}{16\kappa^2}\int dt e^{-\Phi}\tr(\dS^4)\labell{action3}
 \eeqa
which is consistent with the cosmological action \reef{cosm}. The laps function can be inserted in the  above action  by replacing $dt\rightarrow dt/n^3$.

\section{Cosmological reduction at order $\alpha'^2$}

In this section, we are going to show that up to one-dimensional field redefinitions and total derivative terms, the cosmological reduction of the couplings \reef{S2f} can be written in $O(25,25)$-invariant form. 

Since these couplings involve Riemann curvature, $H$, $\nabla H$, $\nabla\phi$ and $\nabla\nabla\phi$, one needs  the cosmological reduction of the Riemann curvature and $H$ which are given in \reef{red1}, and the reduction of $\nabla H$, $\nabla\Phi$, $\nabla\nabla\Phi$, $\nabla G_{ij}$, and $\nabla\nabla G_{ij}$ which are 
\beqa
&&\nabla_l H_{ijk}=-\frac{1}{2}\dB_{jk}\dG_{il}+\frac{1}{2}\dB_{ik}\dG_{jl}-\frac{1}{2}\dB_{ij}\dG_{kl}\,\,;\,\,\nabla_0H_{ij0}=-\frac{1}{2}\dB_j{}^k\dB_{ik}-\frac{1}{2}\dB_i{}^k\dG_{jk}+\ddot{B}_{ij}\nn\\
&&\nabla_0H_{ijk}=0\,\,;\,\, \nabla_k H_{ij0}=0\,\,;\,\,\nabla_0\Phi=\dP\,\,;\,\,\nabla_i\Phi=0\,\,;\,\,\nabla_0\nabla_i\Phi=0\,\,;\,\,\nabla_0\nabla_0\Phi=\ddot{\Phi}\nn\\
&&\nabla_i\nabla_j\Phi=-\frac{1}{2}\dP\dG_{ij}\,\,;\,\,\nabla_0G_{ij}=\dG_{ij}\,\,;\,\,\nabla_iG_{jk}=0\,\,;\,\,\nabla_0\nabla_iG_{jk}=0\,\,;\,\,\nabla_0\nabla_0G_{ij}=\ddot{G}_{ij}\nn\\
&&\nabla_i\nabla_j G_{kl}=-\frac{1}{2}\dG_{kl}\dG_{ij}
\eeqa
Using the above reductions,  one finds the following cosmological reduction for  the action \reef{S2f}:
\beqa
\bS_2^c&=&-\frac{2b_1^2}{\kappa^2}\int dt e^{-\Phi}\Bigg[- \frac{73}{60} \dB_{i}{}^{k} \dB^{ij} \dB_{j}{}^{l} \dB_{k}{}^{m} 
\dB_{l}{}^{n} \dB_{mn} -  \frac{1}{5} \dB_{i}{}^{k} \dB^{ij} 
\dB_{l}{}^{n} \dB^{lm} \dG_{jm} \dG_{kn}\labell{s21}\\&&\qquad\qquad\qquad\qquad -  \frac{39}{10} 
\dB_{i}{}^{k} \dB^{ij} \dB_{j}{}^{l} \dB^{mn} \dG_{km} \dG_{ln} + 
\frac{3}{4} \dB^{ij} \dB^{kl} \dG_{i}{}^{m} \dG_{j}{}^{n} \dG_{km} 
\dG_{ln} \nn\\&&\qquad\qquad\qquad\qquad+ \frac{39}{5} \dB_{i}{}^{k} \dB^{ij} \dB_{j}{}^{l} 
\dB_{k}{}^{m} \dG_{l}{}^{n} \dG_{mn} -  \dB^{ij} \dB^{kl} \dG_{ik} 
\dG_{j}{}^{m} \dG_{l}{}^{n} \dG_{mn} \nn\\&&\qquad\qquad\qquad\qquad+ \frac{1}{2} \dB_{i}{}^{k} 
\dB^{ij} \dG_{j}{}^{l} \dG_{k}{}^{m} \dG_{l}{}^{n} \dG_{mn} -  
\frac{1}{6} \dG_{i}{}^{k} \dG^{ij} \dG_{j}{}^{l} \dG_{k}{}^{m} 
\dG_{l}{}^{n} \dG_{mn}+\cdots\Bigg]\nn
\eeqa
where dots represent the terms which have trace of $\dG$, $\dG\dG$, $\dB\dB$ and $\dB\dB\dG$, or have $\dP,\ddot{\Phi},\ddot{G},\ddot{B}$. All these terms are removable by field redefinitions and total derivative terms. In above action we have also choose the gauge $n=1$.

To use the total derivatives terms and the field redefinition freedom to remove the dots in  \reef{s21}, we  add all total derivative terms  and all field redefinitions at order $\alpha'^2$ with arbitrary coefficients to  \reef{s21}. We add the following total derivative terms:
\beqa
-\frac{2}{\kappa^2}\int dt\frac{d}{dt}(e^{-\Phi}\cI_2)
\eeqa
where  $\cI_2$ is   all possible    terms at five-derivative level with even parity which are constructed from $\dP$, $\dB$, $\dG$, $\ddot{\Phi}$, $\ddot{B}$, $\ddot{G}$,  $\cdots$.  Using the package   "xAct" \cite{Nutma:2013zea}, one finds there are 132 such terms, \ie
\beqa
\cI_2&=&j_1(\dP)^5+\cdots\labell{I2}
\eeqa
where the coefficients  $J_1,\cdots, J_{132}$ are 132 arbitrary parameters.

One can change the field variables in \reef{creduce} as 
\begin{eqnarray}
G_{ij}&\rightarrow &G_{ij}+\alpha' \delta G^{(1)}_{ij}+\alpha'^2 \delta G^{(2)}_{ij}\nn\\
B_{ij}&\rightarrow &B_{ij}+ \alpha'\delta B^{(1)}_{ij}+ \alpha'^2\delta B^{(2)}_{ij}\nn\\
\Phi &\rightarrow &\Phi+ \alpha'\delta\Phi^{(1)}+ \alpha'^2\delta\Phi^{(2)}\nn\\
n &\rightarrow &n+ \alpha' \delta n^{(1)}+ \alpha'^2 \delta n^{(2)}\labell{gbpn2}
\end{eqnarray}
where the first order perturbations   $\delta G^{(1)}_{ij}$, $\delta B^{(1)}_{ij}$, $\delta\Phi^{(1)}, \delta n^{(1)}$ are given in \reef{dG1dB1} and the second order perturbations   $\delta G^{(2)}_{ij}$, $\delta B^{(2)}_{ij}$, $\delta\Phi^{(2)}, \delta n^{(2)}$ are all possible  terms at 4-derivative level constructed from $\dP$, $\dB$, $\dG$, $\ddot{\Phi}$, $\ddot{B}$, $\ddot{G}$, $\cdots$.  The perturbations  $\delta G^{(2)}_{ij}$, $\delta\Phi^{(2)}$, $\delta n^{(2)}$ contain even-parity terms and $\delta B^{(2)}_{ij}$ contains odd-parity terms, \ie
\beqa
\delta n^{(2)}&=&n_1\dB_i{}^j\dB_j{}^k\dB_k{}^l\dB_l{}^i+\cdots\nn\\
\delta \Phi^{(2)}&=&e_1\dB_i{}^j\dB_j{}^k\dB_k{}^l\dB_l{}^i+\cdots\nn\\
\delta G^{(2)}_{ij}&=&d_1\dB_i{}^k\dB_k{}^l\dB_l{}^m\dB_m{}_j+\cdots\nn\\
\delta B^{(2)}_{ij}&=&f_1\dG_i{}^k\dB_k{}_j\dB_l{}^m\dB_m{}^l+\cdots\labell{np2}
\eeqa
The coefficients $n_1,\cdots, n_{52}$, $e_1,\cdots, e_{52}$, $d_1,\cdots, d_{121}$  and $f_1,\cdots, f_{59}$ are arbitrary parameters.  

When the field variables in $S_0^c$  are changed according to the above  field redefinitions, they produce two sets of  couplings at order $\alpha'^2$. One set is produced by the second order perturbations, \ie
\beqa
\delta_2 S_0^c&=&-\frac{2}{\kappa^2}\int dt e^{-\Phi}\Bigg[\delta n^{(2)}\left(-\frac{1}{4}\dB_{ij}\dB^{ij}-\frac{1}{4}\dG_{ij}\dG^{ij}+\dP^2\right)\labell{dS0c1}\\
&&+\delta \Phi^{(2)}\left(-\frac{1}{4}\dB_{ij}\dB^{ij}-\frac{1}{4}\dG_{ij}\dG^{ij}+\dP^2\right)-2\dP\frac{d}{dt}\delta \Phi^{(2)}\nn\\
&&+\delta G^{(2)}_{ij}\left(-\frac{1}{2}\dB_k{}^j\dB^{ki}-\frac{1}{2}\dG_k{}^j\dG^{ki}\right)+\frac{1}{2}\dG^{ij}\frac{d}{dt}\delta G^{(2)}_{ij}+\frac{1}{2}\dB^{ij}\frac{d}{dt}\delta B^{(2)}_{ij}\Bigg]\nn
\eeqa
which are similar to the first order perturbation \reef{dS0c}, and the other set is reproduced by square of the first order perturbations, \ie
\beqa
\frac{1}{2}\delta_1^2 S_0^c&=&-\frac{2}{\kappa^2}\int dt e^{-\Phi}\Bigg[\frac{1}{4} \frac{d}{dt}\delta B^{(1)}_{ij} \frac{d}{dt}\delta B^{(1)}{}^{ij} + \frac{1}{4} 
\frac{d}{dt}\delta G^{(1)}_{ij} \frac{d}{dt}\delta G^{(1)}{}^{ij} -  \frac{d}{dt}\delta \Phi^{(1)} \frac{d}{dt}\delta \Phi^{(1)}\nn\\&& -  \dB^{ij} \frac{d}{dt}\delta B^{(1)}_{i}{}^{k} \delta G^{(1)}_{jk} -  \dG^{ij} \frac{d}{dt}\delta G^{(1)}_{i}{}^{k} \delta G^{(1)}_{jk} + \frac{1}{4} \dB^{ij} \dB^{kl} 
\delta G^{(1)}_{ik} \delta G^{(1)}_{jl} + \frac{1}{4} \dG^{ij} \dG^{kl} 
\delta G^{(1)}_{ik} \delta G^{(1)}_{jl}\nn\\&& + \frac{1}{2} \dB_{i}{}^{k} 
\dB^{ij} \delta G^{(1)}_{j}{}^{l} \delta G^{(1)}_{kl} + \frac{1}{2} 
\dG_{i}{}^{k} \dG^{ij} \delta G^{(1)}_{j}{}^{l} \delta G^{(1)}_{kl} -  
\frac{1}{2} \dB^{ij} \frac{d}{dt}\delta B^{(1)}_{ij} \delta n^{(1)} -  \frac{1}{2} \dG^{ij} \frac{d}{dt}\delta G^{(1)}_{ij} \delta n^{(1)}\nn\\&& + 2 \dP \frac{d}{dt}\delta \Phi^{(1)} 
\delta n^{(1)} + \frac{1}{2} \dB_{i}{}^{k} \dB^{ij} \delta G^{(1)}_{jk} 
\delta n^{(1)} + \frac{1}{2} \dG_{i}{}^{k} \dG^{ij} \delta G^{(1)}_{jk} 
\delta n^{(1)} \nn\\&& +\Big( \frac{1}{4} \dB_{ij} \dB^{ij} + 
\frac{1}{4} \dG_{ij} \dG^{ij}  -  \dP^2\Big) 
\delta n^{(1)}\delta n^{(1)} + \Big(\frac{1}{8} \dB_{ij} \dB^{ij}  + \frac{1}{8} \dG_{ij} \dG^{ij} -  
\frac{1}{2} \dP^2\Big) \delta\Phi^{(1)} \delta \Phi^{(1)} \nn\\&& -  \frac{1}{2} \dG^{ij} \frac{d}{dt}\delta G^{(1)}_{ij} \delta \Phi^{(1)} + 2 \dP \frac{d}{dt}\delta \Phi^{(1)} \delta 
\Phi^{(1)} + \frac{1}{2} \dB_{i}{}^{k} \dB^{ij} \delta G^{(1)}_{jk} 
\delta \Phi^{(1)} + \frac{1}{2} \dG_{i}{}^{k} \dG^{ij} \delta  G^{(1)}_{jk} \delta \Phi^{(1)}\nn\\&& -  \frac{1}{2} 
\dB^{ij} \frac{d}{dt}\delta B^{(1)}_{ij} \delta \Phi^{(1)}+\Big( \frac{1}{4} \dB_{ij} \dB^{ij} + \frac{1}{4} \dG_{ij} \dG^{ij} -  \dP^2\Big) \delta n^{(1)} \delta \Phi^{(1)} \Bigg]\labell{d1}
\eeqa
where the first order perturbations are given in \reef{dG1dB1}.

When the field variables in $O(25,25)$-invariant action \reef{action2} are changed according to the field redefinition \reef{gbpn2}, one also finds   the following couplings at order $\alpha'^2$:
\beqa
\delta_1 S_1^c&=&-\frac{2b_1}{\kappa^2}\int dt e^{-\Phi}\Bigg[- \dB^{ij} \dG_{i}{}^{k} \dG_{k}{}^{l} \frac{d}{dt}\delta B^{(1)}_{jl} + 
\frac{1}{2} \dB_{i}{}^{k} \dB^{ij} \dB_{j}{}^{l} \frac{d}{dt}\delta B^{(1)}_{kl} - 
 \frac{1}{2} \dB^{ij} \dG_{i}{}^{k} \dG_{j}{}^{l} \frac{d}{dt}\delta B^{(1)}_{kl} \nn\\&&
-  \frac{1}{2} \dB^{ij} \dB^{kl} \dG_{ik} \frac{d}{dt}\delta G^{(1)}_{jl} + 
\dB_{i}{}^{k} \dB^{ij} \dG_{j}{}^{l} \frac{d}{dt}\delta G^{(1)}_{kl} + \frac{1}{2} 
\dG_{i}{}^{k} \dG^{ij} \dG_{j}{}^{l} \frac{d}{dt}\delta G^{(1)}_{kl}\nn\\&& -  
\frac{1}{2} \dB^{ij} \dB^{kl} \dG_{i}{}^{m} \dG_{km} \delta 
G^{(1)}_{jl} -  \dB_{i}{}^{k} \dB^{ij} \dG_{j}{}^{l} \dG_{l}{}^{m} \delta 
G^{(1)}_{km} -  \frac{1}{2} \dB_{i}{}^{k} \dB^{ij} \dB_{j}{}^{l} 
\dB_{k}{}^{m} \delta G^{(1)}_{lm} \nn\\&&+ \dB^{ij} \dB^{kl} \dG_{ik} 
\dG_{j}{}^{m} \delta G^{(1)}_{lm} -  \frac{1}{2} \dB_{i}{}^{k} \dB^{ij} 
\dG_{j}{}^{l} \dG_{k}{}^{m} \delta G^{(1)}_{lm} -  \frac{1}{2} 
\dG_{i}{}^{k} \dG^{ij} \dG_{j}{}^{l} \dG_{k}{}^{m} \delta G^{(1)}_{lm}\nn\\&& -  
\frac{3}{8} \dB_{i}{}^{k} \dB^{ij} \dB_{j}{}^{l} \dB_{kl} \delta 
n^{(1)} + \frac{3}{4} \dB^{ij} \dB^{kl} \dG_{ik} \dG_{jl} \delta n^{(1)} -  
\frac{3}{2} \dB_{i}{}^{k} \dB^{ij} \dG_{j}{}^{l} \dG_{kl} \delta 
n^{(1)}\nn\\&& -  \frac{3}{8} \dG_{i}{}^{k} \dG^{ij} \dG_{j}{}^{l} \dG_{kl} 
\delta n^{(1)} -  \frac{1}{8} \dB_{i}{}^{k} \dB^{ij} \dB_{j}{}^{l} 
\dB_{kl} \delta \Phi^{(1)} + \frac{1}{4} \dB^{ij} \dB^{kl} \dG_{ik} 
\dG_{jl} \delta \Phi^{(1)}\nn\\&& -  \frac{1}{2} \dB_{i}{}^{k} \dB^{ij} 
\dG_{j}{}^{l} \dG_{kl} \delta \Phi^{(1)} -  \frac{1}{8} \dG_{i}{}^{k} 
\dG^{ij} \dG_{j}{}^{l} \dG_{kl} \delta \Phi^{(1)}\Bigg]\labell{d2}
\eeqa
where we have used the fact that the lapse function appears in the $O(25,25)$-invariant action \reef{action2} by replacing $dt\rightarrow dt/n^3$. The reason for this is that the action has two metric $G^{00}$ which produce $1/n^4$, and the cosmological reduction of $e^{-2\phi}\sqrt{-G}$ is $ne^{-\Phi}$.

Inserting the first order perturbations \reef{dG1dB1} into \reef{d1}, \reef{d2}, and inserting the arbitrary second order perturbations \reef{np2} into \reef{dS0c1}, 
 one finds the cosmological action \reef{s21} can be written as 
 \beqa
 S_2^c&=&\!\!\bS_2^c+\delta_1S_1^c-\frac{1}{2}\delta_1^2S_0^c+\delta_2S_0^c\,=\,-\frac{2b_1^2}{\kappa^2}\int dt e^{-\Phi}\Bigg[\frac{1}{12} \dB_{i}{}^{k} \dB^{ij} \dB_{j}{}^{l} \dB_{k}{}^{m} 
\dB_{l}{}^{n} \dB_{mn} \nn\\&&+ \frac{1}{4} \dB_{i}{}^{k} \dB^{ij} 
\dB_{l}{}^{n} \dB^{lm} \dG_{jm} \dG_{kn} -  \frac{1}{2} 
\dB_{i}{}^{k} \dB^{ij} \dB_{j}{}^{l} \dB^{mn} \dG_{km} \dG_{ln} + \frac{1}{4} \dB^{ij} \dB^{kl} \dG_{i}{}^{m} \dG_{j}{}^{n} \dG_{km} 
\dG_{ln}\nn\\&& + \frac{1}{2} \dB_{i}{}^{k} \dB^{ij} \dB_{j}{}^{l} 
\dB_{k}{}^{m} \dG_{l}{}^{n} \dG_{mn} -  \frac{1}{2} \dB^{ij} 
\dB^{kl} \dG_{ik} \dG_{j}{}^{m} \dG_{l}{}^{n} \dG_{mn} + \frac{1}{2} 
\dB_{i}{}^{k} \dB^{ij} \dG_{j}{}^{l} \dG_{k}{}^{m} \dG_{l}{}^{n} 
\dG_{mn} \nn\\&&+ \frac{1}{12} \dG_{i}{}^{k} \dG^{ij} \dG_{j}{}^{l} 
\dG_{k}{}^{m} \dG_{l}{}^{n} \dG_{mn}\Bigg]\labell{action21}
 \eeqa
for some specific values for the parameters  $n_1,\cdots, n_{52}$, $e_1,\cdots, e_{52}$, $d_1,\cdots, d_{121}$  and $f_1,\cdots, f_{59}$ and up to some total derivative terms \reef{I2}.  We have chosen the parameters in the second order perturbations \reef{np2} and in the total derivative terms such that  the terms in $S_2^c$ which have trace of $\dG$, $\dG\dG$, $\dB\dB$ and $\dB\dB\dG$, or have $\dP,\ddot{\Phi},\ddot{G},\ddot{B}$ and their higher derivatives, are cancelled. Since we are not interested in studying the couplings at order $\alpha'^3$, we don't  write here the explicit form of the second order field redefinitions and the total derivative terms. 
 
Now using the definition of the generalized metric in \reef{S}, one finds the expression inside the bracket above is $-\tr(\dS^6)/24$. Hence the cosmological reduction of the couplings \reef{S2f} can be written as 
\beqa
 S_2^c&=&\frac{2b_1^2}{24\kappa^2}\int dt e^{-\Phi}\tr(\dS^6)\labell{action41}
 \eeqa
up to some field redefinitions and total derivative terms. It is consistent with the cosmological action \reef{cosm}. This  confirms  the effective action \reef{S2f} that has been found in \cite{Garousi:2019mca}.

\vskip .3 cm

Note added: During the completion of this work, the preprint \cite{Codina:2021cxh} appeared which has some overlaps with the results in this paper.

\vskip .3 cm
{\bf Acknowledgements}:  I would like to thank D.~Marques for discussion.

\end{document}